\documentstyle[aps,prl,multicol,psfig]{revtex}
\begin{document}
\title{Anomalous isotopic effect near the charge-ordering 
quantum criticality}
\author{S. Andergassen, S. Caprara, C. Di Castro, and M. Grilli}
\address{Istituto Nazionale per la Fisica della Materia - Unit\`a di Roma 1,
and Dipartimento di Fisica - Universit\`a di Roma ``La Sapienza'',
Piazzale Aldo Moro 2, I-00185 Roma, Italy}
\maketitle

\begin{abstract}
Within the Hubbard-Holstein model, we evaluate the various crossover lines 
marking the opening of pseudogaps in the cuprates, which, in our scenario,
are ruled by the proximity to a charge-ordering quantum criticality (stripe 
formation). We provide also an analysis of their isotopic dependencies, 
as produced by critical fluctuations. We find no isotopic shift of the 
temperature $T^0$ marked as a reduction of the quasiparticle density of states
in various experiments, and a substantial positive shift of the 
pseudogap-formation temperature $T^*$. We infer that the superconducting
critical temperature $T_c$ has almost no shift in the optimally- and 
overdoped regimes while it has a small negative isotopic shift in the 
underdoped, which increses upon underdoping. We account also for
the possible dynamical nature of the charge-ordering transition, and explain 
in this way the spread of the values of $T^*$ and of its isotopic shift, 
obtained with experimental probes having different characteristic timescales.
\end{abstract}


\begin{multicols}{2}
There are several experimental evidences \cite{TALLON} that the peculiar 
properties of the cuprates, both in the normal and in the superconducting 
phase are controlled by a Quantum Critical Point (QCP), located near the 
optimal doping $\delta=\delta_{opt}$. In this framework the phase diagram of 
the cuprates is naturally partitioned into a (nearly) ordered, a quantum 
critical, and a quantum disordered region corresponding to the under-, 
optimally, and over-doped regions respectively. The ordered region occurs
below a second-order transition line $T_?(\delta)$, which depends on the 
nature of the underlying ordering and, upon increasing $\delta$, ends at 
$T=0$ in a QCP (at $\delta=\delta_c\gtrsim \delta_{opt}$). The above
correspondence between the theoretical and experimental partitioning of the
phase diagram, leads to a close connection \cite{TALLON} between the 
hypothetical $T_?(\delta)$ and the crossover line $T^*(\delta)$, below which 
a pseudogap behavior is observed in NQR, NMR relaxation-rate, XANES, and ARPES 
measurements [see, e.g., Ref. \cite{BILLINGE} for a recent overview on 
${\rm La_{2-\delta}Sr_\delta Cu O_4}$ (LSCO)]. In the 
proximity of quantum criticality, critical fluctuations mediate a singular 
interaction among the quasiparticles which can account for both the violation 
of the Fermi-liquid behavior observed in the normal phase of the cuprates, 
and the strong pairing mechanism leading to high-$T_c$ superconductivity 
\cite{TALLON,AFM,VARMA,CAST}. Various realizations of this 
scenario have been proposed, associated with different 
quantum criticalities, e.g., antiferromagnetic \cite{AFM}, excitonic 
\cite{VARMA}, change in the symmetry of the superconducting order parameter 
\cite{SACH}, or incommensurate charge-density wave \cite{CAST}. 

In the QCP framework, any mechanism shifting the position of the QCP is 
mirrored by corresponding shifts in $T_?(\delta)$ and in the superconducting 
critical line $T_c(\delta)$. In particular, the observation of isotopic 
effects (IE's) on $T_c$ \cite{FRANCK,CRAWFORD} and on $T^*$ 
\cite{RAFFA,WILLIAMS,RUBIO} suggests that a lattice mechanism underlies the 
instability marked by the QCP. We here consider the single-band 
Hubbard-Holstein model as a minimal model to describe the strongly correlated 
electrons coupled to the lattice, giving rise to a QCP for the onset of a 
phonon-induced incommensurate charge ordering (CO) \cite{CAST,PERALI,BTDG}.
This introduces density inhomogeneities on a semimicroscopic scale, and 
corresponds to the onset of stripes \cite{notanomenclatura}, coming from the 
high-doping regime. Thus, in our approach, the critical line $T_?(\delta)$ 
corresponds to the line for CO, $T_{CO}(\delta)$, which in real materials 
can be masked by pair formation and lattice effects. 

In this letter, we first determine the mean-field (m-f) critical line 
$T_{CO}^{(0)}(\delta)$, which we identify with the weak-pseudogap crossover 
line $T^0(\delta)$ [$\gg T^*(\delta)$] observed in Knight-shift, transport, 
and static susceptibility measurements \cite{BILLINGE}, as the incipient
depression of the single-particle density of states (DOS). Indeed, for 
$T<T_{CO}^{(0)}$ one expects the CO fluctuations to become 
substantial, leading to a reduction of the quasiparticle DOS, 
which accounts for the weak-pseudogap behavior.
Then, we investigate the effect of fluctuations near the CO QCP in order to 
determine: a) the fluctuation-corrected critical line $T_{CO}(\delta)$, which 
we relate to the pseudogap crossover line $T^*(\delta)$; b) the IE 
on $T_{CO}(\delta)$, from which we shall also infer the effect on 
$T_c(\delta)$. In particular we shall describe the highly non-trivial effect 
of quantum criticality in determining IE's on $T^*$ and $T_c$, 
which can be strong and weak, respectively, and opposite in sign in the 
underdoped cuprates. We shall also show why both $T^*$ and its isotopic shift 
are observed to be larger in experiments with shorter characteristic 
timescales. The near absence of IE on $T_c$ near and above 
$\delta_{opt}$ will also be naturally accounted for.

{\it --- The model and the mean-field instability ---} 
Our two-dimensional (2D) Hamiltonian is 
\begin{eqnarray}
H & = & -t \sum_{\langle i,j \rangle , \sigma} 
\left( c^\dagger_{i\sigma} c_{j\sigma} + H.c.\right) 
-t' \sum_{\langle \langle i,j \rangle \rangle , \sigma} 
\left( c^\dagger_{i\sigma} c_{j\sigma} + H.c.\right)\nonumber \\ 
& - &\mu_0\sum_{i\sigma} n_{i\sigma}
  +  U\sum_i n_{i\uparrow}n_{i\downarrow}
 + {1 \over 2 } \sum_{{\bf q}\ne 0} V_C({\bf q}) 
\rho_{\bf q} \rho_{-{\bf q}} \nonumber \\
& + & \omega_0 \sum_i a^\dagger_i a_i + g \sum_{i,\sigma}
\left( a^\dagger_i+a_i\right) \left( n_{i\sigma} -\langle n_{i\sigma} \rangle 
\right) 
,\label{HHHam}
\end{eqnarray}
where $c^{(\dagger)}_{i\sigma}$ are the fermion operators, 
$a^{(\dagger)}_{i}$ are the phonon operators, $\langle i,j \rangle$ and 
$\langle \langle i,j \rangle\rangle$ indicate nearest- and 
next-nearest-neighbor sites, coupled by the hopping parameters $t$ and $t'$ 
respectively. The lattice spacing has been set to unity. The chemical potential
$\mu_0$ is coupled to the local electron density 
$n_{i\sigma}\equiv c^\dagger_{i\sigma} c_{i\sigma}$, $U$ is the on-site
Hubbard repulsion, and $\rho_{\bf q} \equiv \sum_{{\bf k},\sigma} 
c^\dagger_{{\bf k}+{\bf q},\sigma}c_{{\bf k},\sigma}$. 
$V_C({\bf q})$ ($\approx V_C/|{\bf q}|$ at small $|{\bf q}|$) is the Coulomb 
interaction between electrons in a 2D plane embedded in the three-dimensional 
(3D) space \cite{CAST,BTDG}. $\omega_0$ is the phonon frequency and $g$ is 
the Holstein electron-phonon coupling. In the limit
$U\to\infty$, this model was solved with a standard slave-boson technique, at 
leading order within a large-N expansion in Refs. \cite{CAST,BTDG}. The most 
relevant result was that a CO instability with a finite wavevector ${\bf q}_c$,
incommensurate with the underlying lattice, was found at $T=0$ when $\delta$ 
is reduced below a critical value $\delta_c^{(0)}$. Besides a small, 
weakly momentum-dependent, residual repulsion, nearby this instability 
the critical charge fluctuations mediate a singular scattering
\begin{equation}
\label{gamma}
\Gamma({\bf q},\omega_n)  \approx  - {V \over \xi_0^{-2} + 
\vert {\mathbf q}-{\mathbf q}_c \vert^2 +\gamma \vert \omega_n \vert},
\end{equation}
of strength $V$, between the quasiparticles. Here $\omega_n$ is a bosonic 
Matsubara frequency, $\gamma \sim t^{-1}$ is a characteristic timescale, and 
$\xi_0^{-2}$ is the m-f inverse square correlation length, which measures
the distance from criticality. Eq. (\ref{gamma}) displays the behavior of a 
Gaussian QCP with a dynamical critical index $z=2$ \cite{notaz2}.

Quite remarkably, the complicated formal structure of the quasiparticle 
scattering, mediated by slave bosons, phonons and by the Coulomb 
interaction, is well represented near criticality by a RPA resummation 
$\Gamma({\bf q},\omega_n)\approx V_{eff}({\bf q})/[1+V_{eff}({\bf q})
\Pi ({\bf q},\omega_n)]$ of an effective static interaction
$V_{eff}({\bf q})={\tilde U}({\bf q})+V_C({\bf q})-\lambda t$ \cite{SEIBOLD}. 
Here $\Pi$ is the fermionic polarization bubble in  a 2D lattice,  
$\lambda\equiv 2g^2/t\omega_0$ is the dimensionless electron-phonon coupling, 
and ${\tilde U}({\bf q})\simeq A+B|{\bf q}|^2$ is the residual short-range 
repulsion between quasiparticles. For the correspondence 
of $V_{eff}({\bf q})$ with the parameters of Eq. (\ref{HHHam}) see Ref. 
\cite{SEIBOLD}. The instability condition, which occurs for reasonable values 
$\lambda \sim 1$, is $1+V_{eff}({\bf q}_c)\Pi({\bf q}_c, \omega=0)=0$.
At $T=0$ this determines ${\bf q}_c$ and the position of the m-f QCP 
$\delta_c^{(0)}$. For realistic parameters we find
${\bf q}_c \approx (\pm 1, 0)$ or $(0,\pm 1)$, and $\delta_c^{(0)}
\approx 0.2$ \cite{CAST,BTDG}. By expanding $1+V_{eff}\Pi$
near the instability at $T=0$, we find
$\xi_0^{-2}\propto \delta-\delta_c^{(0)}$, which gives the Gaussian index 
$\nu=1/2$  for $\xi_0$. The m-f critical line $T_{CO}^{(0)}(\delta)$ (the 
solid-line curve in Fig. 1), which starts from the m-f QCP at $\delta_c^{(0)}$
is obtained by considering the $T$ dependence of the bare 
polarization bubble, which reduces to the simple Fermi-liquid form, 
$\propto T^2$, at low $T$. We identify this m-f transition with the 
experimental crossover line $T^0$, extrapolating to low $T$ at a 
doping, which we identify with our $\delta_c^{(0)}$. From the data reported 
in Ref. \cite{BILLINGE}, for LSCO we estimate $\delta_c^{(0)}\approx 0.22$. 
Then we evaluate $T_{CO}^{(0)}(\delta)$ by taking standard quasiparticle 
(i.e., dressed by the slave bosons) band 
parameters ($t_{qp}=0.2$ eV, $t'_{qp}=-0.05$ eV, $\omega_0=0.07$ eV, leading
to $A=0.2$ eV and $B=0.17$ eV in the residual repulsion $\tilde U$). 
$V_C=0.22$ eV and $g=0.21$ eV are adjusted to match the experimental 
extrapolation of 
$T^0$ with the $T=0$ m-f instability (i.e. $\delta_c^{(0)}$). We point out 
that the agreement between $T_{CO}^{(0)}(\delta)$ and $T^0(\delta)$ at finite 
$T$ is obtained without any further adjustment of the parameters. Similar 
parameters are taken for ${\rm Bi_2Sr_2CaCu_2O_8}$ (Bi2212) to fit the data 
in Refs. \cite{DING,ISHIDA}, with $V_C=0.22$ eV and $g=0.23$ eV.

\begin{figure}
{\psfig{figure=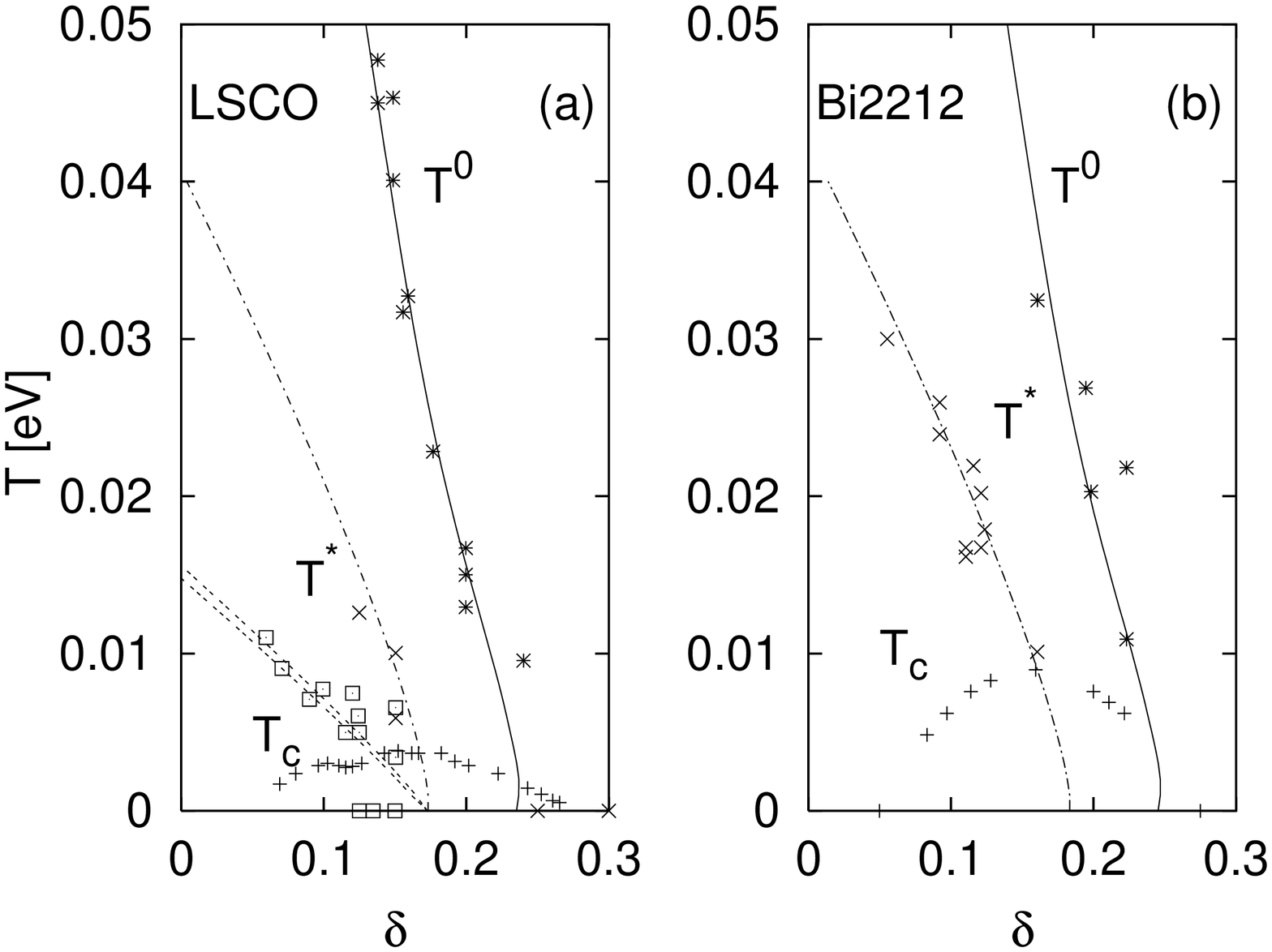,width=8.5cm,angle=-90}}
\end{figure}
\vspace {-0.5truecm}
{\small FIG. 1. 
The phase diagram of the cuprates according to the CO-QCP scenario 
for LSCO (a) and Bi2212 (b). The solid line is the m-f critical line ending 
at $T=0$ in the m-f QCP at $\delta_c^{(0)}$. The lowest dashed line in panel
(a) marks the 3D critical line in the presence of fluctuations, ending in the
QCP at $\delta_c$. We took $\Omega_\perp=1$ meV (see text). The dot-dashed 
line 
in panels (a,b) indicates the ``dynamical instability'' condition (see text)
for $\omega_{probe}=1$ meV. The intermediate dashed line in panel (a)
represents the ``dynamical instability'' condition for 
$\omega_{probe}=1\, \mu$ eV. The experimental points for $T^0$ ($\ast$) and 
for $T^*$ measured with fast ($\times$) and slow ($\Box$) probes for 
LSCO are from Ref.\cite{BILLINGE}, those for Bi2212 are from 
Refs.\cite{DING,ISHIDA}. The experimental critical temperatures $T_c$ are 
also shown  ($+$).
}
\vspace{.3cm}

--- {\it The phase diagram beyond mean-field} ---
The fluctuations shift the m-f QCP and critical line to their actual 
position \cite{HERTZ,MILLIS}. Specifically the fluctuations which mediate the
effective critical interaction, Eq. (\ref{gamma}), can be included in the
polarization bubble via the diagrams of Fig. 2a, leading to corrections beyond
RPA. From the explicit evaluation of these diagrams we find the 
(self-consistent) correction to the mass term $m\equiv \gamma^{-1}\xi^{-2}$ 
of the fluctuation propagator, Eq. (\ref{gamma}), 
\begin{equation}
m=m_0+12uT\sum_{|\omega_n|<\omega_0}\sum_{\bf q}D({\bf q}, \omega_n; m),
\label{mass1}
\end{equation}
where $D=\left(\Omega_{\bf q}+ |\omega_n|\right)^{-1}$ is the 
charge-fluctuation critical propagator, $u\sim V^2/(\gamma^2 t^3)$ is the 
coupling resulting from the two 4-leg vertices represented in Fig. 2b. The 
critical modes have a 2D dispersion 
$\Omega_{\bf q}=\gamma^{-1} |{\bf q}-{\bf q}_c|^2+m$, up to an ultraviolet 
bandwidth cutoff $\Omega_{max}\sim \gamma^{-1}$, resulting from the underlying
lattice. Since we deal with a phonon-driven CO instability, as it also 
follows from the detailed dynamical analysis of Ref. \cite{BTDG}, 
$\omega_0$ appears as the ultraviolet frequency cut-off.  
By introducing the DOS
$N(\Omega)=\sum_{\bf q} \delta(\Omega-\Omega_{\bf q})$ and the
spectral-density representation of $D$, we rewrite Eq. (\ref{mass1}) in the 
form
\begin{eqnarray}
m & = &m_0+{24u \over \pi}\int_{m}^{\Omega_{max}}d\Omega N(\Omega) 
\int_0^\infty d z {z \over z^2 +\Omega^2} \nonumber \\
&\times & 
\left[b(z)+{1 \over 2} - {1\over \pi}{\rm arctg} {z \over \omega_0} \right],
\label{mass2}
\end{eqnarray}
where $b(z)=[\exp{(z/T)}-1]^{-1}$ is the Bose function. At $T=0$, Eq. 
(\ref{mass2}) with $m=0$ leads to a finite shift of the 2D QCP
$\delta_c^{(0)}-\delta_c \propto \omega_0$ (see Fig. 
1). At finite $T$, as a consequence of a constant DOS for the 
2D modes, the integral in Eq. (\ref{mass2}) is logarithmically divergent for 
$m\to 0$, leading to a vanishing of the renormalized $T_{CO}$
(Mermin-Wagner theorem). This divergency is removed by considering the more 
realistic anisotropic 3D character of the critical fluctuations, introducing 
a small energy scale $\Omega_\perp$, below which the mode DOS is 
no longer constant, and displays a 3D square-root behavior 
$N(\Omega<\Omega_\perp) \sim \sqrt{\Omega}$ at criticality. This is enough to 
make the integral in Eq. (\ref{mass2}) convergent and allows to determine
the critical line (Eq. (\ref{mass2}) with $m=0$) in the anisotropic 3D case 
as reported in Fig. 1a (lowest dashed line). The inclusion of fluctuations 
brings the critical line from temperatures of the order of typical electronic 
energies ($T_{CO}^{(0)} \sim t$) down to much lower temperatures $T_{CO}$ of 
the order of the observed $T^*$'s. Indeed, within the CO-QCP scenario, the 
pseudogap arises at $T<T^*$ because the quasiparticles feel an 
increasingly strong attractive interaction by approaching the critical line 
$T_{CO}(\delta)$. In the particle-hole channel, this interaction can produce 
a gap due to the incipient CO. At the same time in the particle-particle 
channel, the strong attraction can lead to pair formation even in the absence 
of phase coherence. Therefore $T^*(\delta)$ closely tracks the underlying 
transition line $T_{CO}(\delta)$.

On the other hand, the spread in the measured values of $T^*$ depending
on the experimental probe (see, e.g., Ref. \cite{BILLINGE}), indicates that 
the CO instability may be ``dynamical'', $\Omega_\perp$ being smaller than 
(or comparable to) typical frequencies of the experimental probes,
$\omega_{probe}$. In this case the self-consistency condition (\ref{mass2}) 
includes $\omega_{probe}$ as the infrared cut-off in 
the integral over $\Omega$, $m$ being replaced by $\omega_{probe}$. This 
determines the doping and temperature dependence of the dynamical instability 
lines. Two examples are reported in Fig. 1 for $\omega_{probe}=1$meV 
(dot-dashed line if Fig. 1a,b), as in typical neutron scattering experiments, 
and $\omega_{probe}=1~\mu$eV (second dashed line from bottom in Fig. 1a), as 
in static experiments (NQR, NMR). The corresponding experimental data for 
$T^*$ in LSCO are also reported for comparison. We determine the coupling $V$ 
between the charge fluctuations and the quasiparticles, which is the only 
parameter for which an a priori estimate is difficult, by imposing that the 
fluctuation-corrected QCP is located at the $T=0$ extrapolation of the 
$T^*(\delta)$ curves. We used $\gamma=0.7$ eV$^{-1}$ and $\gamma=0.4$ eV$^{-1}$
for LSCO and Bi2212 respectively, and $V=0.54$ eV for both. It is 
worth noticing that, similarly to the case of $T_{CO}^{(0)}(\delta)$, the 
agreement between the calculated $T_{CO}(\delta)$ and the experimental points 
$T^*(\delta)$ is obtained without further adjustable parameters. We also 
notice that, contrary to the shift of $\delta_c$ at $T=0$, the slope of the 
curve $T_{CO}(\delta)$ is weakly dependent on $\omega_0$.
\vspace{-0.5truecm}
\begin{figure}
{\hspace{1 truecm}\psfig{figure=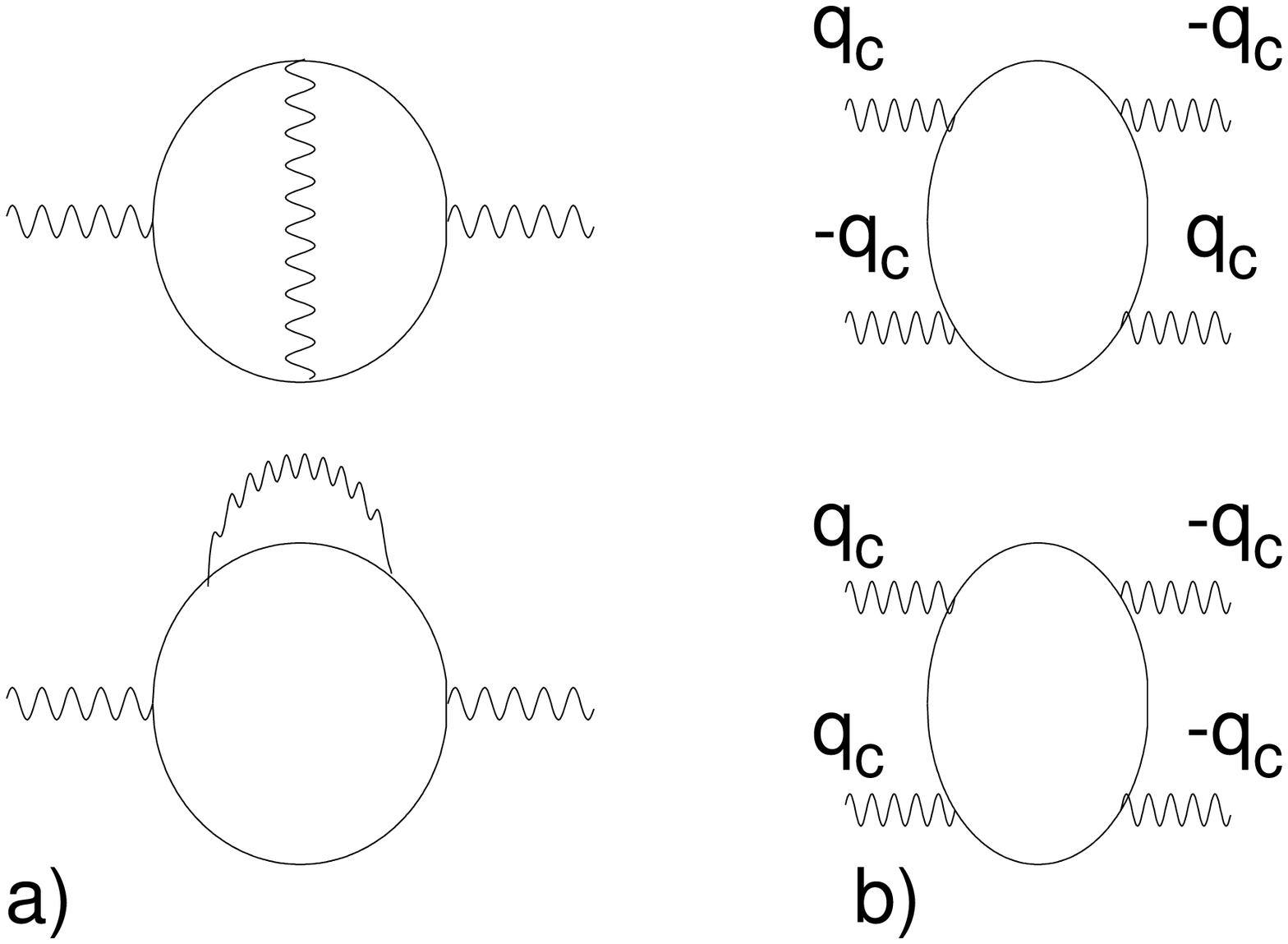,width=5.5cm,angle=-90}}
\end{figure}
\vspace {-0.5 truecm}
{\small FIG. 2. (a) 
The two (vertex and selfenergy) corrections to the fermionic 
bubbles (solid line) due to critical charge fluctuations (wavy line). 
(b) Effective 4-leg vertices for the critical charge fluctuation field (wavy 
line) resulting from the integration over the fermion loops (solid line).  
}

--- {\it Novel isotopic effects} ---
The m-f weak-pseudogap crossover temperature $T^{(0)}_{CO}\sim T^0$ is 
determined by $\lambda$ only. Therefore, it does not depend on $\omega_0$ and 
is {\it not} expected to display any isotopic dependence. On the other 
hand, quantities determined by the fluctuations crucially involve $\omega_0$. 
New physical effects can then arise in the isotopic substitutions. Since 
$\omega_0$ decreases by increasing the ionic mass, the corrections to 
$\delta_c^{(0)}$ and to $T_{CO}^{(0)}(\delta)$ become smaller and the 
fluctuation-corrected quantities $\delta_c$ and $T_{CO}$ shift to higher 
doping, closer to the m-f values, as shown in Fig. 3, where we report the 
line $T_{CO}(\delta)$, calculated via Eq. (\ref{mass2}), with parameters to 
fit the $T^*$ data of LSCO (see above), together with its isotopic shift 
calculated for $^{16}O\to \, ^{18}O$ substitution (i.e., for a five percent 
reduction of $\omega_0$). Correspondingly we expect that the portion of the 
curve $T_c(\delta)$ on the left of $T^0(\delta)$ is rigidly translated along 
the horizontal axis. The rational behind this translation is that the whole 
physics of these materials is essentially determined by the proximity to the 
critical line $T_{CO}(\delta)$ and to the related QCP. On the other hand, on 
the right of the m-f critical line $T_{CO}^{(0)}(\delta)\sim T^0(\delta)$ the 
fluctuations are small and the physical processes at $T>T^0(\delta)$ are 
captured by the m-f description, where $\lambda$ (not $\omega_0$) is relevant.
As a consequence $T^0(\delta)$ and the portion of $T_c(\delta)$ near and 
above $T^0(\delta)$ are not expected to be shifted by IE's.

In the underdoped region, there are two evident consequences of the isotopic 
shift, which becomes more substantial: The shift upon reducing $\omega_0$ is 
negative (as usual) in $T_c$, but, contrary to standard theories based on CO 
pseudogap \cite{EREMIN}, it is {\it positive} in $T_{CO}\sim T^*$. Moreover, 
when the slope of $T_{CO}(\delta)$ is large, a rather small isotopic shift in 
$\delta_c$ can result in a substantial shift in $T_{CO}\sim T^*$. The steeper 
is $T^*$, the larger is the IE. On the other hand, since 
the curve $T_c(\delta)$ is rather flat, particularly in the optimal and 
moderately under-doped regimes, the expected IE on $T_c$ in these
compounds is small in agreement with long-standing experiments
\cite{FRANCK,CRAWFORD}. This large difference in the IE for
$T_c$ and $T^*$, $\left(\Delta T_c/\Delta M\right)/\left(\Delta T^*/\Delta 
M\right)\ll 1$, is indeed experimentally observed in ${\rm HoBa_2Cu_4O_8}$ 
(HBCO-124), and reported in Ref. \cite{RUBIO}, where it is also noticed that 
there is a ``striking similarity between isotopic substitution and 
underdoping with respect to both $T_c$ and $T^*$''. Although we are not aware 
of any systematic analysis of the doping dependencies of $T_c$, $T^*$ and 
their isotopic shifts in HBCO-124, this observation finds its natural 
interpretation within our QCP scenario, where the isotopic substitution 
produces a shift of the QCP and is therefore nearly equivalent to underdoping.
\begin{figure}
{\hspace{0.5 truecm}{\psfig{figure=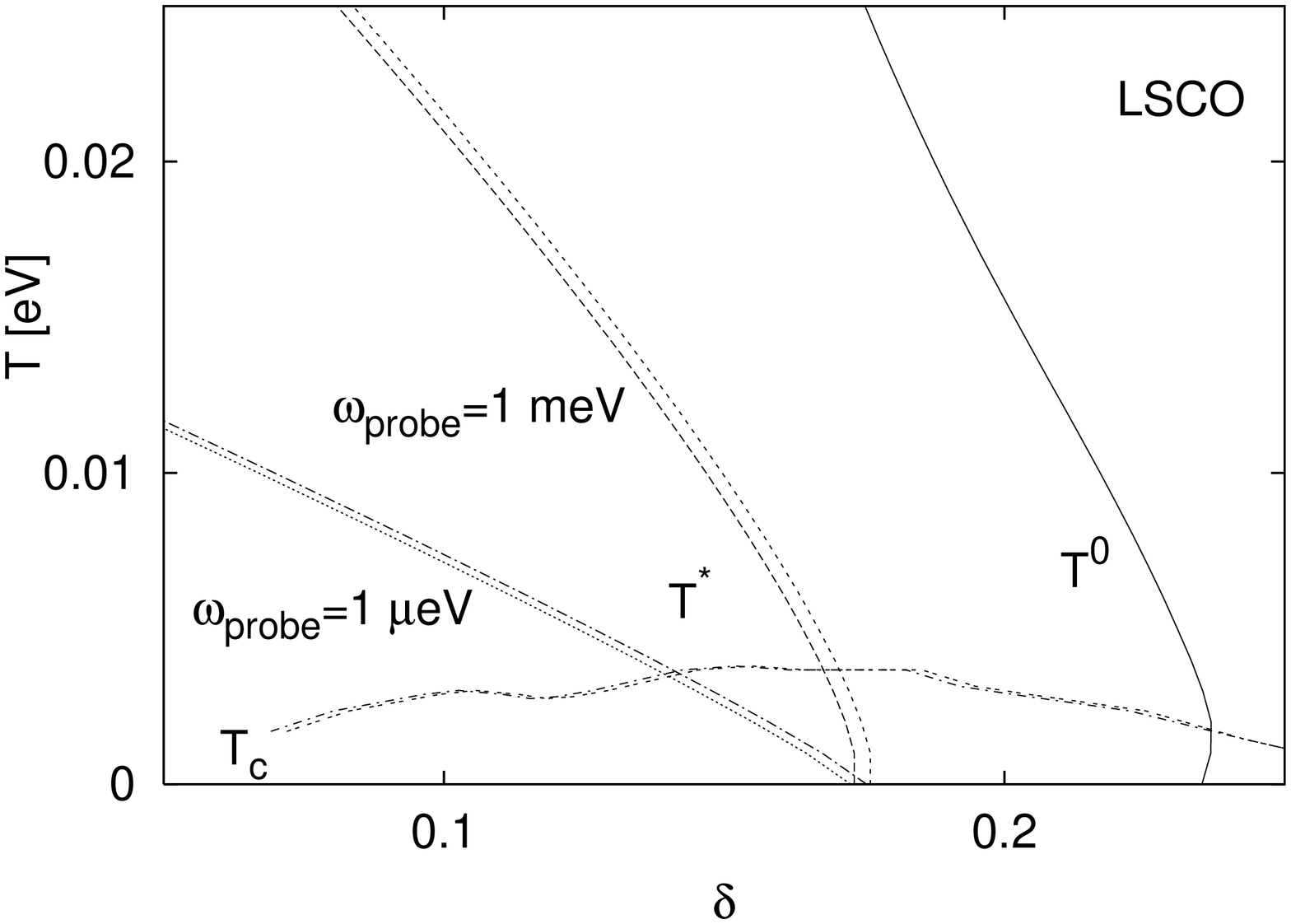,width=7.cm,angle=-90}}}
\end{figure}
{\small FIG. 3. 
Calculated effect of the isotopic change ${\rm ^{16}O \to ^{18}O}$ 
(i.e. $\omega_0=70$ meV and $\omega_0'=66$ meV) on $T^*$ in LSCO, both for a 
fast- and a slow-probe measurement. The inferred shift on $T_c$ is also 
reported}
\vspace{.5cm}

The slope of $T_{CO}(\delta)$ increases by increasing $\omega_{probe}$,  
while the QCP is unshifted. Therefore,  the IE on 
$T_{CO}\sim T^*$ is enhanced, and we have the general trend that faster 
probes should detect a larger IE on $T^*$, since the effect of
fluctuation diminishes. Although we cannot 
account for the near-absent or negative IE on $T^*$ within the 
almost static probes in ${\rm YBa_2Cu_4O_8}$ \cite{RAFFA,WILLIAMS}, this 
general trend is in qualitative agreement with a much stronger effect 
observed in the isostructural HBCO-124 with fast neutron scattering 
\cite{RUBIO}. Indeed, this fast-probe experiment should be represented by 
the curve $T_{CO}(\delta)$ corresponding to $\omega_{probe}=1$ meV 
in Fig. 3. The huge IE on $T^*$ observed in HBCO-124 suggests 
that the curve $T_{CO}(\delta)$ in this system is steeper than in LSCO.

Upon underdoping, while $T_c$ decreases, its IE increses in agreement
with a general experimental trend \cite{FRANCK}. However,
in the strongly underdoped materials, where a substantial IE in 
the penetration depth \cite{HOFER} and in the x-ray absorption \cite{LANZARA} 
have been observed, our results can provide only qualitative indications, as 
other effects (magnetic, polaronic, lattice-pinning) not included in 
(\ref{HHHam}) become relevant. For this doping regime, a 
different (additional) interpretation of the IE in terms of a 
superconductor-to-insulator QCP has been recently proposed \cite{SCHNEIDER}.

{\em Acknowledgments.} We acknowledge stimulating discussions
with C. Castellani and K. A. M\"uller.

\vspace {-0.5 truecm}

\end{multicols}

\begin{references}
\vspace {-1.75 truecm}
\bibitem{TALLON}
        For a summary of this evidences see, e.g., the Introduction of 
        C. Castellani, C. Di Castro, and M. Grilli, Z. f\"ur Physik
        {\bf 103}, 137 (1997);
        J. L. Tallon, {\it et al.}, Phys. Stat. Sol. (b) 215, 531 (1999);
        J. L. Tallon, and J. W. Loram, Physica C {\bf 349}, 53 (2001).
\bibitem{BILLINGE} M. Gutmann, E. S. Bo\v zin, and S. J. L. Billinge,
	cond-mat/0009141.
\bibitem{AFM} A. J. Millis, H. Monien, and D. Pines, Phys. Rev. B {\bf 42},
	167 (1990); P. Monthoux, A. V. Balatsky, and D. Pines, Phys. Rev. B 
	{\bf 46}, 14 803 (1992); S. Sachdev and J. Ye, Phys. Rev. Lett. 
        {\bf 69}, 2411 (1992).
\bibitem{VARMA} C. M. Varma, Phys. Rev. Lett. {\bf 83}, 3538 (1999).
\bibitem{CAST} C. Castellani, C. Di Castro, and M. Grilli, Phys. Rev. Lett.
        {\bf 75}, 4650 (1995).
\bibitem{SACH} M. Vojta, Y. Zhang, and S. Sachdev,  Phys. Rev. Lett. 
	{\bf 85}, 4940 (2000).
\bibitem{FRANCK} J. P. Franck, in {\it Physical Properties of High
	Temperature Superconductors IV}, edited by D. M. Ginsberg
	(World Scientific, Singapore, 1994), p. 189.
\bibitem{CRAWFORD} M. K. Crawford, {\it et al.}, Phys. Rev. B {\bf 41}, 282
	(1990).
\bibitem{RAFFA} F. Raffa, {\it et al.}, Phys. Rev. Lett. {\bf 81}, 5912 (1998).
\bibitem{WILLIAMS} G. V. M. Williams, {\it et al.}, Phys. Rev. B
	{\bf 61}, R9257 (2000).
\bibitem{RUBIO} D. Rubio Temprano, {\it et al.}, Phys. Rev. Lett.
	{\bf 84}, 1990 (2000).
\bibitem{PERALI} A. Perali, {\it et al.}, Phys. Rev. B {\bf 54}, 16 216 (1996).
\bibitem{BTDG} F. Becca, {\it et al.}, Phys. Rev. B {\bf 54}, 12 443 (1996).
\bibitem{notanomenclatura}In our framework the CO
	instability is a second-order transition leading to a
	gradual increase of the charge modulation. It is only when
	one enters deeply inside the (locally) ordered phase that
	anharmonic distortions of the charge profile may arise
	from the enhanced interactions with the spin and the lattice
	degrees of freedom. This may lead to the stripe formation, as
	revealed in neutron scattering,	see J. M. Tranquada, {\sl et al.},
	Nature {\bf 375}, 561 (1995).
\bibitem{notaz2} This feature is the natural consequence of the damping of 
	the charge fluctuations due to the creation of particle-hole pairs at 
	arbitrarily low energy when ${\mathbf q}_c$ joins two ``hot'' points 
	on the Fermi surface. 
\bibitem{SEIBOLD} G. Seibold, {\it et al.}, Eur. Phys. J. B {\bf 13},
	87 (2000).
\bibitem{DING} H. Ding, {\it et al.}, J. Phys. Chem. Solids {\bf 59},
	1888 (1998).
\bibitem{ISHIDA} K. Ishida, {\it et al.}, Phys. Rev. B {\bf 58},
	R5960 (1998).
\bibitem{HERTZ} J. A. Hertz, Phys. Rev. B {\bf 14}, 1165 (1976).
\bibitem{MILLIS} A. J. Millis, Phys. Rev. B {\bf 48}, 7183 (1993).
\bibitem{EREMIN}I. Eremin, {\it et al.}, Phys. Rev. B {\bf 56},
	11 305 (1997).
\bibitem{LANZARA} A. Lanzara, {\it et al.}, J. Phys.: Condens. Matter {\bf 11},
	L541 (1999).
\bibitem{HOFER} J. Hofer, {\it et al.}, Phys. Rev. Lett. {\bf 84}, 4192 (2000).
\bibitem{SCHNEIDER} T. Schneider and H. Keller, cond-mat/0011381.
\end{references}
\end{document}